# Structural sensitivity of the spin Hall magnetoresistance in antiferromagnetic thin films

Andrew Ross[1,2], Romain Lebrun[1], Camilo Ulloa[3], Daniel A. Grave[4], Asaf Kay[4], Lorenzo Baldrati[1], Florian Kronast[5], Sergio Valencia[5], Avner Rothschild[4], Mathias Kläui[1,2,6]

1. *Institut für Physik, Johannes Gutenberg-Universität Mainz, D-55099, Mainz, Germany*
2. *Graduate School of Excellence Materials Science in Mainz (MAINZ), Staudingerweg 9, D-55128, Mainz, Germany*
3. *Institute for Theoretical Physics, Utrecht University, Princetonplein 5, 3584 CC Utrecht, The Netherlands.*
4. *Department of Materials Science and Engineering, Technion-Israel Institute of Technology, Haifa 32000, Israel.*
5. *Helmholtz-Zentrum Berlin für Materialien und Energie, Albert-Einstein-Strasse 15, D-12489 Berlin, Germany.*
6. *Center for Quantum Spintronics, Department of Physics, Norwegian University of Science and Technology, Trondheim, Norway.*

**Abstract**

**Reading the magnetic state of antiferromagnetic (AFM) thin films is key for AFM spintronic devices. We investigate the underlying physics behind the spin Hall magnetoresistance (SMR) of bilayers of platinum and insulating AFM hematite (α-$Fe_2O_3$) and find an SMR efficiency of up to 0.1%, comparable to ferromagnetic based structures. To understand the observed complex SMR field dependence, we analyse the effect of misalignments of the magnetic axis that arise during growth of thin films, by electrical measurements and direct magnetic imaging, and find that a small deviation can result in significant signatures in the SMR response. This highlights the care that must be taken when interpreting SMR measurements on AFM spin textures.**

**Introduction**

With spin dynamics in the terahertz regime and a robustness to external perturbations from magnetic fields, antiferromagnetic spintronics seeks to make use of this exciting class of materials for future spintronic devices [1,2]. A pure spin current can be generated in insulating antiferromagnets (AFMI) by several means [3–6]. However, regardless of the excitation mechanism, the detection of such a pure spin current relies on the inverse spin-Hall effect (SHE) in a heavy metal (HM) layer in contact with the AFMI layer. For spin-Hall magnetoresistance (SMR), a charge current $J_C$ flowing in the HM leads to a transverse spin current that flows towards the AFMI/HM interface where it is either absorbed or reflected by the AFMI. This then leads to a modulation of the resistance of the HM as the angle between $J_C$ and the magnetic order is changed. One can thus theoretically determine the orientation of the antiferromagnetic order parameter, the Néel vector ***n***, for both collinear [7–9] and some non-collinear [10] magnetic structures by means of electrical measurements without the need for complex synchrotron-based measurements [11]. There have however been conflicting reports about SMR measurements in antiferromagnets, exhibiting so-called positive SMR [12,13] and negative SMR [7,8] on AFMs as well as the role of the antiferromagnetic symmetry and the interface quality.

Whilst the SMR is expected to depend on the orientation of ***n***, significant parasitic contributions can arise from both ordinary magnetoresistance (OMR) [7,14] and the net magnetic moment of thin film canted antiferromagnets [13] and ferrimagnets [10]. The SMR has also been investigated in $Cr_2O_3$, which has a locally uncompensated interface that has no noticeable impact on the angular dependent SMR signals [15,16]. On the theory side, AFMs have been largely treated analogously to ferromagnets although a ferromagnetic interface is uncompensated, leading to real and imaginary components of the spin mixing conductance (SMC) [17].

In this letter, we make use of the insulating antiferromagnet hematite, $\alpha$-$Fe_2O_3$, which exhibits a phase transition from an easy-plane AFM to an easy-axis (EA) AFM, known as the Morin transition at $T_M$ [18]. We study the SMR for different antiferromagnetic symmetries without changing the interface by varying the temperature at which the SMR is recorded, finding distinctly different SMR behaviours. Below $T_M$, the shape of the SMR cannot be explained using conventional SMR theory. Taking into account the interfacial symmetry then proves critical in the spin transmission. In particular, for our growth direction, the out-of-plane strain on the crystal lattice leads to a small but finite deviation of the magnetic axis from the film normal, as we confirm from direct magnetic imaging. This deviation then heavily influences the resulting SMR signal of easy-axis antiferromagnets highlighting that great care must be taken when analysing AFM-SMR responses.

**Main Text**

In order to investigate the SMR, epitaxial (0001) oriented 100 nm thin films of hematite were deposited on like-wise orientated sapphire ($Al_2O_3$) via pulsed laser deposition from a stoichiometric $Fe_2O_3$ target at 800°C [11,19]. Hall bars were defined by electron beam lithography and the subsequent deposition of 7 nm platinum by DC sputtering without additional interfacial treatments. We do not perform additional interfacial treatments due to the propensity of transition metal oxides to restructure under etching [20]. In the case of hematite, this would be into other (potentially ferrimagnetic) iron oxides [21]. By making use of magnetic dichroism, we do not observe any evidence of such ferrimagnetic oxides at the interface [22]. A charge current $J_C$ is passed through the Pt Hall bar along the *x*-axis indicated in Fig. 1(a) which produces a spin accumulation $\mu_s$ at the interface, polarised along *y* whilst a longitudinal voltage $V_L$ is detected, from which we calculate a resistance $R_L$.

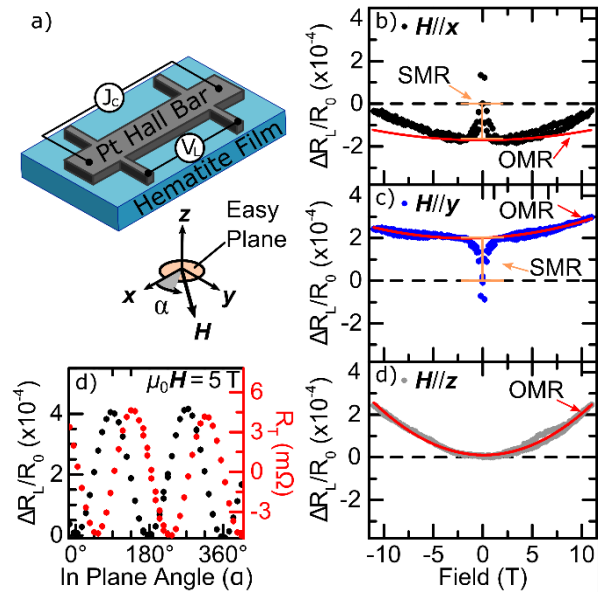

*Fig. 1. (a) Schematic of the measurement geometry employed of a Pt Hall bar atop hematite films. The charge current and longitudinal voltage contacts are indicated. The sample xy-plane is the antiferromagnetic easy plane (indicated in orange) above the Morin transition temperature. Normalized longitudinal resistance ($\Delta R_L = R_L - R_0$, where $R_0$ is the zero-field resistance) of a Pt Hall bar atop (0001) orientated hematite at 300 K in the easy plane antiferromagnetic phase. The resistance is measured as a function of the magnetic field applied along the (b) x-axis, (c) y-axis and (d) z-axis. The red line represent the ordinary magnetoresistance (OMR). (e) Normalized longitudinal (black) and transverse (red) resistance for an in-plane rotation of the magnetic field $\mu_0 H = 5$ T. ($\Delta R_L = R_L - R_{\alpha=0°}$). Error bars, where visible, represent the standard deviation of the measurement points.*

We first investigate the SMR response at room temperature calculated at the change in resistance with respect to the zero-field resistance, normalised to the zero-field resistance ($\Delta R_L/R_0$). At this temperature the *xy*-plane is the antiferromagnetic easy-plane. Like orthoferrites [23], hematite has an antisymmetric exchange interaction directed along the (0001) direction that leads to a canting of **n** [24], generating a net magnetic moment **m**⊥**n** [18,25], where the role **m** plays in the SMR is currently unclear [10,13,26]. In the absence of a magnetic field $H$, $n$ demonstrates a three-fold degeneracy that leads to an observable magnetic domain structure [22,27–29]. $H$ applied within the easy-plane, parallel ($H//x$) or perpendicular ($H//y$) to J$_c$, will break this degeneracy and $n$ will rotate to lie perpendicular to $H$. The strength of the in-plane anisotropy defines a critical field H$_{MD}$ required to produce this complete rotation [7,30]. Indeed, for $H//x$ we observe initially a decrease of R$_L$ up to H$_{MD}$ = 0.6 T (Fig. 1b). At $|H|$>H$_{MD}$ the magnetic signal should then saturate, given that AFM-SMR is dependent on the $y$ component of $n$, i.e. $n_y$ [7,8,26,30,31]. However, we observe a sustained increase with increasing magnetic field. If $H$ is now directed along $y$, (Fig. 1c) there is a steep increase of R$_L$ that plateaus at the same field value H$_{MD}$ as before. Above H$_{MD}$ there is again a parabolic background. The decrease (increase) of the SMR is related to the change between the zero- and finite field orientations of $n$ and $\mu_s$ where for $H//x$, the final state is $n||\mu_s$ (for $H//y$, the final state is $n\perp \mu_s$) and $n_y$ increases (decreases) [26,31].

When the magnetic field is applied out of the plane, along $z$, $n$ is already perpendicular to $H$ and there is no preferred orientation of $n$ within the plane. There is however a parabolic change of R$_L$ with magnetic field (Fig 1d). This parabolic increase in resistance, present in all field directions on top of the magnetic contributions, stems from the ordinary magnetoresistance (OMR) of the Pt itself and is unrelated to $n$ [7,14,32]. This is highlighted for all three uniaxial scans by a red parabolic fit made using Kohler's rule with appropriate values to the high field data (see Fig. 1b-d) [33]. We note that the experimental curve measured in Fig. 1a deviates quantitatively from the expected OMR response at high magnetic fields. Considering that the magnetic field lies perpendicular to the Néel vector, this will induce a canting in the direction of $H$, reducing the component of $n$ parallel to $\mu_s$ at high magnetic fields and may lead to this deviation [26].

Although the behaviour of the SMR for a field along a single direction indicates that $n$ dominates the transport response, it does not exclude the possibility of $m$ playing a role as it has been reported to contribute strongly to the SMR in thin films of a canted antiferromagnet [13]. To identify the governing factor for the SMR, we therefore rotate $H$ in the $xy$-plane through an angle α (Fig. 1e). Here, the longitudinal SMR response is calculated with respect to a field parallel to the charge current $\Delta R_L =$

$R_L - R_{\alpha=0°}$ and normalised by the zero-field resistance of the device. In this plane, the SMR signal for both the longitudinal $R_L$ and transverse $R_T$ resistances shows a behaviour that can be modelled by a $\sin^2\alpha$ relationship indicative of negative SMR and is thus dominated by the orientation of **n**, even in the presence of the spontaneous net moment. The magnitude of the SMR response between the uniaxial measurements (Fig. 1b)-c)) which are relative to the zero-field resistance of the devices, and the angular dependent longitudinal resistance in Fig. 1e) which is determined as the change in the SMR between **H//x** and **H//y** are consistent with one another. Due to the existence of a magnetic domain structure in the absence of a magnetic field in our films [22], the SMR response of antiferromagnets cannot be unambiguously resolved from angular dependent measurements below the critical magnetic fields alone.

Next we check the effect of the antiferromagnetic symmetries, for which the surface sensitive nature of SMR is well suited [10,26]. Below $T_M$, the antiferromagnetic structure has an EA structure and **n** lies along the *z*-axis. For **H⊥EA** (**H//x** and **H//y**), **n** undergoes a transition unique to hematite at some critical field $H_c^\perp$ [34,35], smoothly rotating perpendicular to both the EA and **H**. Considering first the case for **H//x**, **n** rotates and the final state above $H_c^\perp$ is that of **n**//**y**, decreasing the expected SMR. Alternatively, we can apply **H//y**, which also induces a rotation of **n** in the plane parallel to $J_c$ and the SMR ratio is expected to remain constant [17,36]. Finally, **H//z** induces a spin-flop at $H_c^{||}$ and **n** reorients from *z* to lie within the *xy* plane [35].

We investigate the SMR at 175 K, below $T_M$, for different orientations of the magnetic field relative to $J_c$. Starting from **H//x** in Fig 2a, $R_L$ begins to decrease as expected until it reaches $\mu_0 H_c^\perp = 2.5\ T$ and saturates at $3\times10^{-4}$, a higher amplitude than seen above $T_M$. As **H** continues to increase, there are no further changes to the magnetic state and the parabolic nature can again be attributed to the OMR [7,14,32]. We then investigate the SMR for **H//y** where our expectation is to observe only the OMR contribution. However, surprisingly we observe a significant increase of $R_L$ up to the same magnetic field as **H//x**, which must then be related to the orientation of **n**. Before we attempt to explain this signal, we briefly discuss **H//z**, where the resistance begins to decrease with increasing magnetic field due to a small field misalignment and **n** begins to rotate. There is a dominating parabolic effect from the OMR, which masks the smaller effect of the spin-flop field, $\mu_0 H_c^\parallel = 6.5$ T indicated in Fig. 2c for the increasing magnetic field. As the magnetic field is lowered, a hysteretic field dependence appears, where the critical field is shifted slightly due to the additional, stabilising magnetoelastic energy of the spin-flop state above $H_c^\parallel$ leading to different values for the increasing and decreasing field branch [37]. There is an asymmetry in the size of the hysteresis for a negative and positive magnetic field, possibly due to an asymmetry in the magnetostriction of hematite when the magnetic field is not perfectly along the easy axis [37]. If a magnetic field is rotated in the *xy*-plane through an angle α, we again resolve a sin²α dependence characteristic of negative SMR, just as above $T_M$ in Fig. 1e, a feature that arises due to the antisymmetric exchange interaction that enables the rotation of ***n*** at $H_c^\perp$.

To investigate the origin of the signal for ***H//y***, we note that, when we discuss the relative field directions, these are relative to the charge current of the Hall bar. Given the (0001) orientation of the crystal growth, the low temperature AFM structure is expected to lead to an easy-axis parallel to the geometric ***z*** axis (perpendicular to the film plane), which coincides with the (0001) axis of the $Al_2O_3$ substrates. Measurements of the crystalline structure of our films by x-ray diffraction reveal a hematite peak with a FWHM of 0.538±0.009° [22]. This is not unexpected given the 5% lattice mismatch between hematite and sapphire (5.038 Å vs 4.785 Å) resulting in strained growth that relaxes as the films get thicker. However, this could also lead to a slight deviation of the hematite (0001) axis from the film normal [22]. To investigate the effect that a small deviation from the geometric ***z*** axis has on the SMR response, we develop a simple model of hematite (see [22]). Starting first from the naïve assumption of the easy-axis perfectly coincident with the out of plane direction, we reproduce the expected response for ***H//x*** and ***H//y***. We then model the response as a function of angle for a deviation from ***z*** of the EA in both the *xz* plane and the *yz* plane. For a deviation in the *yz* plane, the impact on both field directions is profound, whilst the effect of a deviation in the *xz* plane is minimal. For a deviation between these two extremes, the effect on the SMR will similarly be between these two extremes. Comparing these theoretical expectations with our experimental results, it is clear that a small deviation of ***n*** from the EA gives rise to the experimentally observed SMR response for ***H//y***. It is unlikely that the deviation in our films is confined to a single plane as it is for our simulation. Instead, it likely presents a range of deviations across the average of the Hall bar, and thus the SMR response is the average of all these deviations. To confirm that this model is quantitatively consistent with the thin film magnetic structure, we performed x-ray magnetic linear dichroism – photoemission electron microscopy (XMLD-PEEM) imaging of the same thin film samples used for the electrical measurements oriented as (0001) and compare the magnetic contrast to ($1\bar{1}02$) oriented films grown under the same conditions (see Supplementary Ref. [22]). In the case of the (0001) oriented films, we observe a magnetic contrast, which likely arises from a distribution in the c-axis directions of the films, even in the easy-axis phase, where the 180° nature of the AFM domains should lead to no contrast between them. However, the only impact of this domain structure on the SMR response is the emerging hysteresis in Fig. 2c [22]. In single crystal antiferromagnets, the domain structure is not considered [26] and we find here that the presence of large domains in thin films does not lead to an improved SMR response [22]. Such an effect is not noticeable when measuring the SMR response of an easy plane AFM given the nature of the anisotropy and that the SMR is sensitive to the net change of the projection $\mathbf{n}\cdot\mathbf{\mu}_s$. Only by moving to an easy-axis AFM can this lead to the anomalous signals observed here. Previous measurements of the SMR on the easy-axis AFM $Cr_2O_3$ do not report similar effects [9,16]. However, for the (0001) growth orientation, this effect was only noticeable for $\mathbf{H}\perp$EA, a configuration in $Cr_2O_3$ that does not lead to a spin reorientation and thus does not affect the relative alignment of $\mathbf{n}$. We also note that we do not see a similar effect for ($1\bar{1}02$) oriented films, where the c-axis of the hematite lies parallel to the substrate c-axis and has a projection in-plane [22].

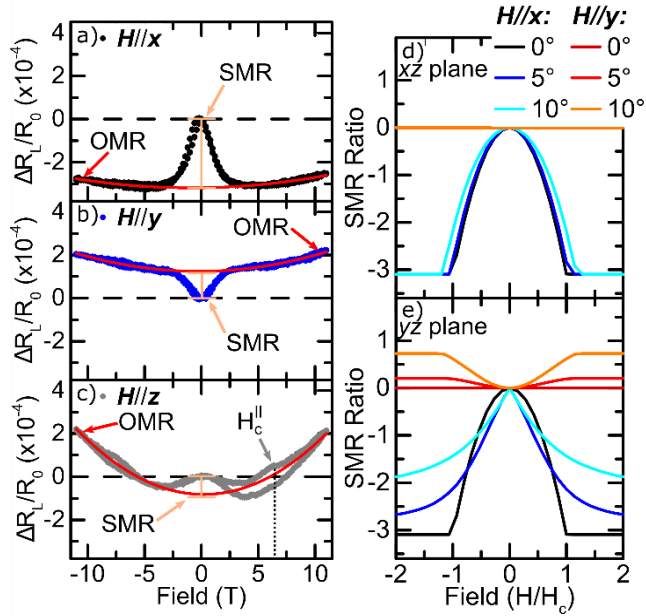

*Fig. 2 Longitudinal resistance of Pt/hematite at 175 K in the easy-axis phase for a magnetic field applied along the (a) x-axis (b) y-axis and (c) z-axis. $\Delta R_L = R_L - R_0$ where $R_0$ is the zero-field resistance. The red line indicates the parabolic contribution from the ordinary magnetoresistance. Error bars where visible represent the standard deviation of the data point. We indicate in c) the critical magnetic field $H_c^{\parallel}$ for an increasing magnetic field. (d)-(e) Expected effect on the SMR ratio for a misalignment of the easy-axis in the xz plane (d) or the yz plane (e) for a magnetic field applied in plane along either **x** or **y**. The critical field of the magnetic field induced 2$^{nd}$ order transition of the Neel vector is $H_c = (2JK_z - D^2)/D$, where J is the strength of the exchange interaction, $K_z$ is the uniaxial anisotropy, and D is the strength of the DMI exchange interaction. This transition describes a rotation of the Neel vector parallel to the easy-axis to a state perpendicular both to the easy-axis and the magnetic field.*

Finally, we turn our attention to the temperature dependence of the SMR. In Fig. 3a the temperature dependence of the SMR for ***H//x*** at select temperatures is shown. The arrows indicate the critical field for rotation of ***n*** at either $H_c^{\perp}$ or $H_{MD}$. We extract the critical fields from the saturation point for $H_c^{\perp}$ and plot it alongside the temperature dependence for $H_c^{\parallel}$, extracted from the SMR for ***H//z***, in Fig. 3b. With decreasing temperature, these field increase, in line with bulk measurements of the critical fields [26,35], however, to our knowledge, the temperature dependence of $H_c^{\perp}$ for thin film hematite has not been reported. The strong temperature dependence of the magnetic anisotropies of hematite determine the critical fields $H_c^{\parallel}$ and $H_c^{\perp}$ as well as $T_M$, where they drop to zero [35]. The temperature dependence of the anisotropy is expected to scale by $H_c \propto \sqrt{T_M - T}$ following a typical temperature dependence of a second-order Landau phase transition [26,28]. We then use this functional form to fit the critical fields of Fig. 3b, indicated by the dashed lines, providing an estimate of $T_M$ ~195 K – 230 K. Through the Morin transition, we see a low value of the SMR ratio for ***H//x*** (Fig. 3c) that decreases as we move away from the transition, the anisotropies begin to increase, and the thermal fluctuations of the transition are suppressed by the lowering temperature [8,28]. The spurious contribution for ***H//y*** (Fig. 3c) experiences a drop of 50% across the transiton, before adopting a behaviour that does not depend on the temperature. This drop across the transition can be explained by the absence of the SMR contribution from the domain redistribution at $H_{MD}$ with the persistent contribution coming from the previously discussed tilt in the easy-axis relative to the lab frame. The increase in the absolute value of the SMR ratio is contrary to the decrease with temperature observed in Ref [32], attributed to anisotropic magnetoresistance (AMR). This contribution to the SMR was only visible for thin Pt layers 2 nm thick, whilst the OMR of thicker films masked this effect [32]. Given the thickness of our Pt (7 nm) and the large OMR contribution, alongside the increase with lower temperatures (Fig. 3c), any AMR contribution to our signal is negligible [32]. Using the data of Fig. 3c, we can estimate a lower limit on the real part of the effective spin mixing conductance $G^{\uparrow\downarrow}$ for our Pt/α-Fe$_2$O$_3$ in the two antiferromagnetic phases [22]. Above the Morin transition in the easy plane phase, we find $G^{\uparrow\downarrow}$ ~ 3.37x10$^{14}$ Ω$^{-1}$m$^{-2}$, a value comparable to other easy plane antiferromagnets [7]. At lower temperatures, we observe a change in the value of $G^{\uparrow\downarrow}$ from $G^{\uparrow\downarrow}$ ~ 4.60x10$^{14}$ Ω$^{-1}$m$^{-2}$ just below the Morin transition to $G^{\uparrow\downarrow}$ ~ 8.20 x10$^{14}$ Ω$^{-1}$m$^{-2}$ at 25 K. Although the temperature dependence of $G^{\uparrow\downarrow}$ is not clear [38–40], we find a continuous decrease of $G^{\uparrow\downarrow}$ across the phase transition from easy plane to easy axis.

From SQUID magnetometry (see [22] and Fig.3d), the Morin transition of the 100 nm films used here can be measured in order to confirm the value of $T_M$ from electrical measurements. If the films are thinner than this, the Morin transition is heavily suppressed or cannot be resolved [41]. We observe that, when cooled in an external field, the Morin transition of our films is visible from the drop in magnetic moment to effectively zero above the background substrate contributions (Fig 3d). The transition takes place over a relatively broad temperature range but below 200 K, the transition is complete and the films are in the purely easy-axis phase. The value of $T_M$ given by SQUID is then comparable not only to synchrotron measurements on similar films [11] but also with the value extracted from our electrical measurements. We note that in other SMR studies on thinner films of hematite [32,42], the films have not demonstrated a Morin transition and remain in the easy-plane phase. It is then only by virtue of having a Morin transition are we able to observe the impact of the crystal growth. SQUID magnetometry can also be used to investigate the magnitude of $H_c^{\parallel}$ and $H_c^{\perp}$, however thicker films are required in order to have sufficient signal to noise in the first derivative of the signal [43]. This further highlights the practicality of SMR in investigating the magnetic anisotropies of AFMs without the need for relying on volume effects.

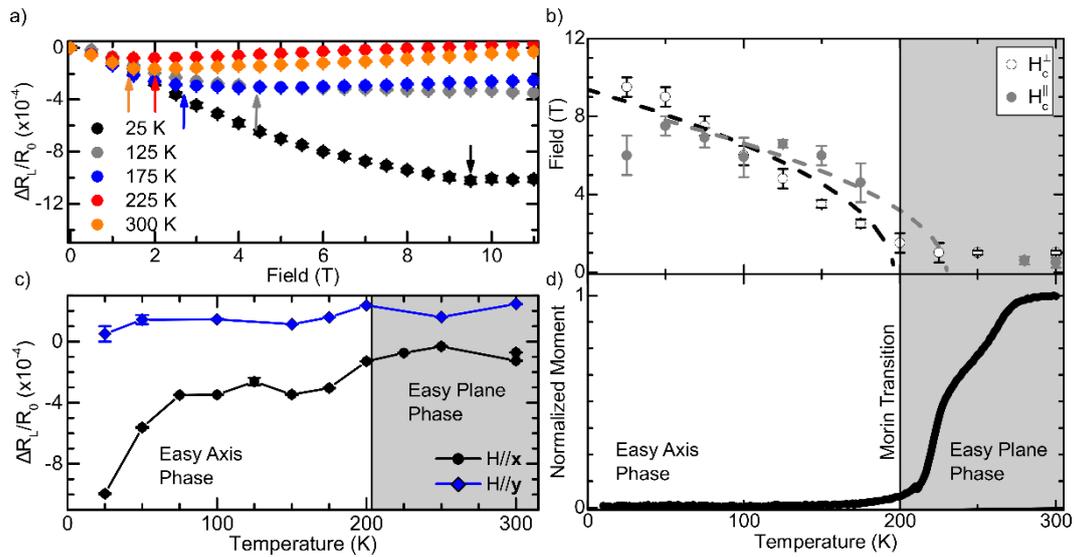

Fig 3. (a) Temperature dependence of the normalised longitudinal resistance for a magnetic field **H//x**. The arrows indicate the critical magnetic field $H_c^{\perp}$ for each temperature. (b) Temperature dependence of the critical fields for **H//x (H//y)** (black) and **H//z** (grey). (c) Temperature dependence of the SMR ratio for **H//x** (black) and **H//y** (blue). (d) Normalised magnetic susceptibility of 100 nm thick (0001) orientated hematite for a field parallel to the c-axis. The Morin transition is indicated.

In conclusion, the spin Hall magnetoresistance technique is shown to be an excellent tool for investigating antiferromagnets to understand the effects of anisotropies, symmetries and surface symmetry breaking. The surface sensitivity nature of the technique allows for the extraction of the Morin transition even for thin films of hematite where conventionally used bulk-sensitive techniques fail. We find distinctly different behaviours above and below the Morin transition. By probing the SMR for different directions of the applied magnetic field, we observe an SMR response that cannot be described by assuming a perfect crystallographic orientation of the film. By considering the distribution of growth crystallites that accompanies thin film deposition, we demonstrate that a small misalignment between the magnetic easy-axis and the external geometric system can have large consequences for the SMR response and the interpretation of the underlying magnetic symmetries. These conclusions are confirmed by direct imaging of the domain structure.

**Acknowledgements**

A.R. and M.K. acknowledge support from the Graduate School of Excellence Materials Science in Mainz (DFG/GSC 266). This work was supported by the Max Planck Graduate Center with the Johannes Gutenberg-Universität Mainz (MPGC). A. R., R. L. and M.K. acknowledge support from the DFG project number 423441604. R.L. acknowledges the European Union's Horizon 2020 research and innovation programme under the Marie Skłodowska-Curie grant agreement FAST number 752195. All authors from Mainz also acknowledge support from both MaHoJeRo (DAAD Spintronics network, project number 57334897), SPIN+X (DFG SFB TRR 173, project A01) and KAUST (OSR-2019-CRG8-4048.2). D.A.G., A.K. and Av.R. acknowledge support from the European Research Council under the European Union's Seventh Framework programme (FP/200702013) / ERC (Grant Agreement No. 617516). D.A.G. acknowledges support from The Center for Absorption in Science, Ministry of Immigrant Absorption, State of Israel. The work including the Mainz-Utrecht collaboration was also supported by the Research Council of Norway through its Centres of Excellence funding scheme, project number 262633 'QuSpin'. C.U. acknowledges support from the Nederlandse Organisatie voor Wetenschappelijk Onderzoek (NWO). L.B. acknowledges the European Union's Horizon 2020 research and innovation programme under the Marie Skłodowska-Curie Grant Agreement ARTES number 793159. Magnetic domain imaging measurements were carried out at the PEEM instrument of the UE49-PGMa beamline at Helmholtz-Zentrum Berlin für Materialien und Energie. We thankfully acknowledge the financial support from HZB under the proposal number: 182-07628 ST.

**Author contributions**

R.L. and M.K. proposed and supervised the project. A.R. performed the transport experiments. A.R. patterned with R.L. the samples. D. G, A. K and Av. R grew and optimized the films. A.R and R.L. analyzed the data with input from M.K. C.U. developed the theoretical model with input from A.R. L.B., R.L., F.K. and S.V. performed the magnetic imaging and analysis. A. R wrote the paper with R.L. and M.K. All authors commented on the manuscript.

Supporting Information

1) X-ray Diffraction Characterization

The crystallographic structure of our films can be investigated by performing x-ray diffraction, using the known (0001) peak of the $Al_2O_3$ substrate for alignment. In Supplementary Fig. S1a, we show a $2\theta$ measurement of a typical α-$Fe_2O_3$ films across the (0001) peaks of both film and substrate, where both are labelled appropriately. We fit a Gaussian distribution (red line) to the α-$Fe_2O_3$ peak and find a full width half maximum (FWHM) of 0.538° and a peak value of $2\theta$ at 39.337°. The $R^2$ value of the Gaussian fit is 0.89. To the left of the main film peak, we observe a small shoulder, indicating regions of the film with a variation in the out-of-plane axis. By fixing the angle of the incident x-rays at $2\theta = 39.337°$, we can perform a rocking curve around this angle. We show this in Supplementary Fig. S1b along with a Gaussian fit in red. From this rocking curve, we find the peak angle to be located at 19.697°. This value differs from the expected value (19.668°) taken from Supplementary Fig. S1a, indicating that the film and substrate [0001] axes are not parallel, and a finite angle exists between them. Aligning to the film peak gives the same result. We can calculate the effective grain size by making use of Scherrer's formula, where we find a lower limit of 17.4 nm. This broader peak may also be related to the formation of domains within each crystallographic grain that are dominated by the magnetoelastic interaction. These domains are visible in magnetic imaging, performed later."

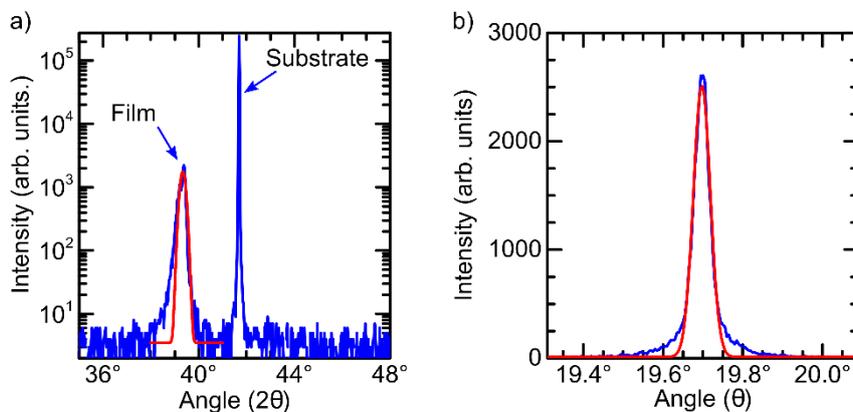

*Supplementary Fig. S1. a) X-ray diffraction measurement of 100 nm thick, (0001) α-$Fe_2O_3$ film on a $Al_2O_3$ substrate across the (0001) peak of both. The α-$Fe_2O_3$ peak is fit by a Gaussian (red line). b) Rocking curve of the α-$Fe_2O_3$ (0001) peak where the incident angle of the x-rays is fixed at $2\theta = 39.337°$. The rocking curve is fit by a Gaussian (red line).*

2) SQUID Magnetometry

In order to investigate the Morin temperature of the hematite thin films, superconducting quantum interference magnetometry (SQUID) was used. At the Morin temperature, the weak moment caused by the additional Dzyaloshinskii-Moriya interaction disappears as the Néel vector rotates. This is then characterized by a drop in the magnetic susceptibility of the samples. The films were first warmed to 400 K, where a large magnetic field of 3 T was applied perpendicular to the easy-axis. This field is in order to saturate the weak moment in the direction of the field, within the easy-plane. The magnetic field is then reduced to a probing field of 50 mT and the sample is cooled in this field at a rate of 2 K/min. A temperature independent, constant background is subtracted to account for the diamagnetic contribution from the sapphire substrate. The magnetic susceptibility is shown in Figure 3d) of the main text and the Morin transition is clearly visible from the drop of the signal to around zero. We label 200 K as the Morin transition where the spin structure has completely rotated. The transition takes place over a relatively broad temperature range unlike in bulk hematite but is an effect seen previously in thin films of hematite.

3) Theoretical Model on the spin Hall magnetoresistance in hematite thin films

Here, we will calculate the spin Hall magnetoresistance (SMR) of a thin film of hematite. Defining the lab frame as (*x,y,z*) where the charge current flows along *x*, we study the effect on the SMR of a small misalignment between the antiferromagnetic easy-axis and *z*.

2.1) Hematite Model
We model the hematite films with a two-spin model described by the Hamiltonian,

$$\mathcal{H} = J\mathbf{S_1} \cdot \mathbf{S_2} - \mathcal{D} \cdot \mathbf{S_1} \times \mathbf{S_2} - \frac{K_z}{2}[(\mathbf{S_1} \cdot \mathbf{e})^2 + (\mathbf{S_2} \cdot \mathbf{e})^2] - \mathbf{H} \cdot (\mathbf{S_1} + \mathbf{S_2}), \qquad (S1)$$

where $\mathcal{J}$ is the exchange interaction, $\mathcal{D} = \mathcal{D}\hat{z}$ is the DMI interaction, $K_z$ is the anisotropy directed along the easy-axis $\mathbf{e}$ and $\mathbf{H}$ is the magnetic field. In the following, we consider the easy axis to be misaligned with *z*. We quantify the misalignment using spherical coordinates $\mathbf{e} = (\sin\theta\cos\phi, \sin\theta\sin\phi, \cos\theta)$. In such a way we consider different degrees of misalignment of the easy-axis quantified by the azimuthal angle θ both in the planes *xz* and *yz*. The spins follow the Landau-Lifschitz-Gilbert equation (LLG) given by,

$$\dot{\mathbf{S}}_i = -\gamma \mathbf{S}_i \times \mathbf{H}_i^{eff} + \alpha \mathbf{S}_i \times \dot{\mathbf{S}}_i, \qquad (S2)$$

where α is the damping parameter and,

$$H_i^{eff} = -\frac{1}{\hbar\gamma}\frac{\delta\mathcal{H}}{\delta S_i} \qquad (S3)$$

Is the effective field for the $i^{th}$ spin. We will solve these equations numerically and write the steady state in terms of the Néel and magnetization order parameters, respectively $\mathbf{n}=1/2(\mathbf{S_1} - \mathbf{S_2})$ and $\mathbf{m}=1/2(\mathbf{S_1} + \mathbf{S_2})$.

2.2) Antiferromagnetic Spin Hall Magnetoresistance
The equations describing the spin hall effect in the Pt Hall bar are given by,

$$j_i = \frac{\sigma}{q}\partial_i\mu - \frac{\sigma'}{2q}\epsilon_{ijk}\partial_j\mu_k, \qquad (S4.1)$$

$$j_{i,j}^s = -\frac{\sigma}{2q}\partial_i\mu_j - \frac{\sigma'}{q}\epsilon_{ijk}\partial_k\mu, \qquad (S4.2)$$

Where $\sigma$ is the conductivity of the platinum, $j_i$ is charge current and $j_{i,j}^s$ is the spin current polarised along the $j^{th}$. The interfacial spin current will then be,

$$j_s = \frac{g_{n,r}^{(\uparrow\downarrow)}}{4\pi} n \times (\mu \times n) + \frac{g_{m,r}^{(\uparrow\downarrow)}}{4\pi} m \times (\mu \times m) + \frac{g_{m,i}^{(\uparrow\downarrow)}}{4\pi} \mu \times m \qquad (S5)$$

We assume that a spin accumulation $\vec{\mu}$ builds up in the platinum bar that follows a diffusion equation given by,

$$\partial_z^2 \vec{\mu} = \vec{\mu}/l_s^2, \tag{S6}$$

Where $l_s$ is the spin diffusion length of the Pt. Equation (S6) has a general solution given by,
$$\vec{\mu}_s(z) = \boldsymbol{\mu}_- e^{-z/l_s} + \boldsymbol{\mu}_+ e^{z/l_s} \tag{S7}$$

Where $\boldsymbol{\mu}_+$ and $\boldsymbol{\mu}_-$ are constants determined by the boundary conditions. The boundary conditions for the spin current in the Pt flowing in the $z$ direction are,
$$j_y^s(z=0) = 0 \tag{S8}$$
while for the other boundary the interfacial current is given by Eq.(S5) evaluated in $z = t$.

Using equation (S7) and its boundary conditions together with Eqs. (S4) we solve for the charge current along the $x$ direction, finding

$$\langle j^q \rangle = \sigma E + e^{t/l_s}\left(e^{t/2l_s} - e^{-t/2l_s}\right)^2 \frac{\sigma'}{qt} \hat{z} \times \mu_+ - \sigma\theta_{SH}^2 \frac{l_s}{t}\left(e^{t/l_s} - 1\right)\hat{z} \times (\hat{z} \times E), \tag{S9}$$

where $j_0 = \sigma E$ is the unperturbed charge current and the constant $\boldsymbol{\mu}_+$ is a function of $\boldsymbol{n}, \boldsymbol{m}$, and $\boldsymbol{E}$ and is computed numerically using Eq.(S5) and the results of the LLG simulation. Meanwhile, $\theta_{SH}$ is the spin Hall angle and $t$ is the thickness of the platinum. We define the resistivity tensor $\rho$ from Eq.(S9) as $\boldsymbol{E} = \rho < \boldsymbol{j^q} >$ and focus on the magnetic field dependence of $\rho$.

We numerically solve Equation S2 with the following physical parameters: $\mathcal{J}/\mu_B = 1000$ T, $\mathcal{D}/\mu_B = 0.65$ T, $K_z/\mu_B = 30$ mT, $\alpha = 0.001$ where $\mu_B$ is the Bohr magneton. We analyze the effect of an external magnetic field on the resistivity tensor in two cases: (i) a magnetic field parallel to the charge current and (ii) a magnetic field perpendicular to the charge current for misalignments of $\boldsymbol{e}$ with respect to $z$. The amplitude was taken in the range H=-2H$_c$ → 2H$_c$ where H$_c$ = $\frac{2\mathcal{J}K_z^2 - \mathcal{D}}{\mathcal{D}}$ the critical field for the second order transition going from $\boldsymbol{n}\|\boldsymbol{e}$ to $\boldsymbol{n}\perp$ to both $\boldsymbol{H}$ and $\boldsymbol{e}$. The results are shown for select angles in the main text Figs. 2c and 2d.

### 4. Spin Mixing Conductance
We calculate the effective spin mixing conductance (SMC) $G^{\uparrow\downarrow}$ according to the formulation provided by Chen *et al* [1] for a magnetic field parallel to $\boldsymbol{x}$. We focus on this orientation given the dependence of the SMC on the measured value of the SMR, where $\boldsymbol{H}$//$\boldsymbol{y}$ (Fig. 3c of the main text) shows minimal temperature dependence compared to $\boldsymbol{H}$//$\boldsymbol{x}$. For $\boldsymbol{H}$//$\boldsymbol{z}$, the dominant OMR contribution makes it challenging to extract the true contribution from the SMR. We use a spin diffusion length and spin Hall angle for our Pt of 1.5 nm and 0.01 respectively [2]. The conductivity of Pt was extracted from our experimental data in the absence of a magnetic field. We show in Supplementary Fig. 2 a comparison between the SMR value for $\boldsymbol{H}$//$\boldsymbol{x}$ and the calculated effective SMC.

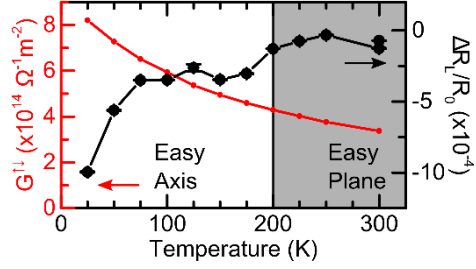

*Supplementary Fig. 2. Calculated effective spin mixing conductance (red) as a function of temperature for a magnetic field applied parallel to the charge current. Comparison curve to the experimental values of the SMR (black) as a function of temperature for the same configuration. The Morin transition temperature obtained from SQUID is marked (Fig. 3d of main text).*

We extract an absolute value of the effective SMC at 300 K of $G^{\uparrow\downarrow} \sim 3.37 \times 10^{14}\, \Omega^{-1} m^{-2}$ and a value that increases with decreasing temperature from $G^{\uparrow\downarrow} \sim 4.60 \times 10^{14}\, \Omega^{-1} m^{-2}$ just below the Morin transition to $G^{\uparrow\downarrow} \sim 8.20 \times 10^{14}\, \Omega^{-1} m^{-2}$ at 25 K. This change may originate from a temperature dependence of other parameters rather than representing a temperature dependence of the spin mixing conductance itself [3–5]. Although the conductivity used in the calculation was temperature dependent, we assumed a constant value of the spin Hall angle and the spin diffusion length. These values are comparable to other reported values of the SMC for AFM/HM bilayers [6,7] and YIG/Pt heterostructures [8].

5. Spin Hall Magnetoresistance in ($1\bar{1}02$) Oriented Hematite Films

For comparison, the spin Hall magnetoresistance has also been investigated for 500 nm ($1\bar{1}02$) oriented hematite films, grown on $Al_2O_3$ substrates with the same orientation by pulsed laser deposition. To allow for a comparison, we have as for the (0001) films discussed in the main text, also patterned Pt Hall bars by lithographic means and the deposition of 7 nm Pt in an Ar atmosphere (Supplementary Fig. 3a). Due to the orientation of growth, the crystallographic c-axis, and therefore the low temperature easy-axis, lies at an angle of 33° to the sample plane. The Hall bar is oriented such that the charge current $J_c$ is passed parallel to the in-plane projection of the easy-axis (Supplementary Fig. 3a). To identify the Morin transition temperature of these films, we perform temperature dependent measurements of the magnetization for a probing magnetic field applied perpendicular and parallel to the c-axis (Supplementary Fig. 3b). The magnetization is measured for both a cooling (-2 K/min) and warming curve (+5 K/min). The Morin transition is visible for the magnetic field perpendicular to the c-axis as the net moment present in the easy-plane phase reduces with decreasing temperature. We label $T_M \sim 200$ K as the point when the films have adopted an easy-axis nature across the majority of the film volume. Just as for the (0001) films, the Morin transition temperature is reduced from the bulk value of 260 K.

We now turn our attention to investigating the SMR in these films. The longitudinal resistance is measured as a magnetic field **H** is swept perpendicular to $J_c$. This places **H** perpendicular to the easy-axis. Given that the SMR depends on the component of the Néel vector perpendicular to $J_c$, i.e. ***n·μ_s***, this orientation should be insensitive to the reorientation of the Néel vector that can occur in the easy-axis phase [9]. This is confirmed by the measurements just below $T_M$ in Supplementary Fig. 3c. If the c-axis was misaligned to the substrate, as in the (0001) films of the main text, then we would see a signal for this configuration of **H** with respect to $J_c$ (see Fig. 2b of the main text). As we increase the temperature, the films adopt an easy-plane nature where ***n*** rotates 90°. This easy-plane has degenerate orientations of ***n*** in the absence of a magnetic field, and thus a finite component of ***n*** will lie parallel to the interfacial spin accumulation ***μ_s***. The application of the magnetic field will then lead ***n*** to reorient within the easy-plane perpendicular to **H** above a certain magnetic field. This will reduce the projection

$n \cdot \mu_s$ leading to an increase in $\Delta R_L$. This behavior is seen in Supplementary Fig. 3c for two temperatures, 230 K within the Morin transition and at 270 K, above the Morin transition.

The SMR in Supplementary Fig. 3c is one order of magnitude smaller than that of the (0001) oriented films in Fig. 1 and Fig. 2 of the main text. This reduction indicates that this crystal orientation has resulted in a poorer Pt/α-Fe$_2$O$_3$ interface than in the films of the main text. We also observe no significant contribution from the ordinary magnetoresistance that produced the parabolic background in Fig 1b-1d and Fig. 2a-2c of the main text. This could be due to several factors such as the quality of the Pt on this orientation of hematite leading to differences in the grain size of the Pt or variations in the Pt deposition, which were performed at different times for the two film orientations.

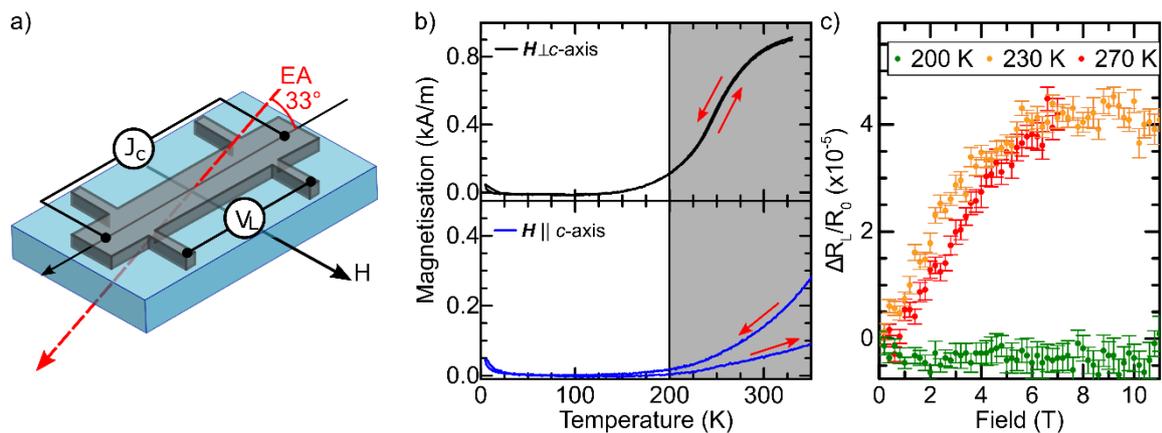

*Supplementary Fig. 3. a) Schematic of a Pt Hall bar atop a 500 nm ($1\bar{1}02$) oriented hematite film. The charge current $J_c$ passes parallel to the in-plane projection of the easy-axis (EA). b) SQUID magnetometry measurements of the magnetization for a magnetic field applied perpendicular (top) or parallel (bottom) to the c-axis. The remnant magnetization of the easy-plane phase decreases with temperature due to the Morin transition to the easy-axis phase. The red arrows indicate cooling and warming curves. c) Change in longitudinal resistance as a function of magnetic field for temperatures above and below the Morin transition temperature. The error bars represent the standard deviation of each data point.*

6. Impact of Antiferromagnetic Domain Structures on the Spin Hall Magnetoresistance
To further elucidate the deviation of the easy-axis from the substrate normal by making use of x-ray magnetic linear dichroism-photoemission electron microscopy (XMLD-PEEM) technique to image the antiferromagnetic domains. These images have been acquired at the Fe L$_2$ edge in 2 energy mode with a single polarization at $hv_1 = 720.24\ eV$ and $hv_2 = 721.79\ eV$, calculating the contrast, after normalizing and correcting for the image drift, from the two images as: $XMLD = \frac{hv_1 - hv_2}{hv_1 + hv_2}$. The beam shines x-rays on the sample inclined at an angle of 16° and we use both linear horizontal (LH) out-of-plane and linear vertical (LV) in-plane polarizations. We removed a parabolic background to get rid of the non-uniform illumination resulting from the dispersion of the beam energy. We show in Supplementary Fig. 4 XMLD-PEEM images for (0001) oriented hematite, capped by 2 nm Pt to mitigate charging effects, at 100 K, i.e. below the Morin transition temperature. If the easy-axis was parallel to the surface normal, then the 180° nature of easy-axis antiferromagnetic domains would not give rise to magnetic contrast between neighboring domains. However, as clearly seen in Supplementary Fig. 4, the antiferromagnetic domain structure can be resolved originating from a distribution of the antiferromagnetic axis with respect to the incoming beam axis. this is further emphasized by comparing incident x-rays polarized in-plane (Supplementary Fig. 4b) with those polarized mostly out-of-plane (Supplementary Fig. 4a). The out-of-plane contrast is clearly stronger, however the in-plane contrast shown in Supplementary Fig. 4b indicates that the Néel vector is rotated away from the surface normal.

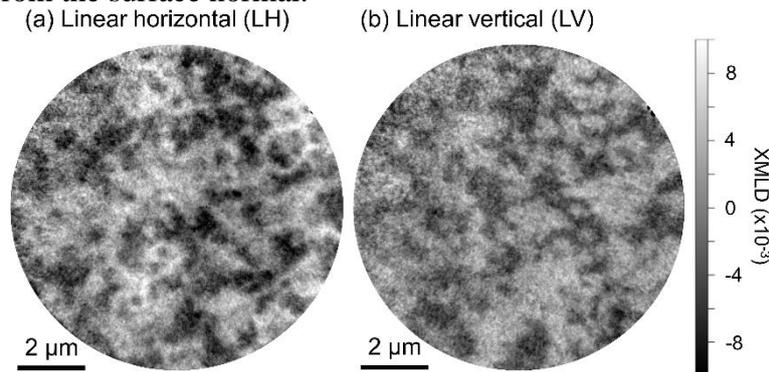

Supplementary Fig. 4. X-ray magnetic linear dichroism photoemission electron microscopy (XMLD-PEEM) images of (0001) oriented hematite capped by 2 nm Pt. a) PEEM image for linear horizontal polarization of X-rays providing out-of-plane contrast. b) PEEM image for linear vertical polarization of X-rays providing in-plane contrast. The contrast in both originates from a deviation of the Néel vector from the surface normal by varying angles in each domain.

We show in Supplementary Fig. 5 XMLD-PEEM imaging of ($1\bar{1}02$) oriented hematite films capped by 2 nm Pt to prevent charging effects at 100 K in the easy-axis phase. The antiferromagnetic domains are clearly visible by the large regions of single contrast typical of an easy-axis antiferromagnet. These large domains are separated by clear domain walls as the Néel vector rotates 180°. When we focus on the in-plane component of the Néel vector, shown in Supplementary Fig. 5b) we see that there is no correlation to the domain structure visible in Supplementary Fig. 5a). This supports the transport measurements in Supplementary Fig. 3c, where we concluded that the easy-axis was collinear with the substrate c-axis. We notice that in Supplementary Fig. 5a) a closed domain wall exists in the upper right quadrant, a feature that may be of interest to the wider spintronics community. Focusing on this feature closer, we observe a circular domain wall with a central core (Supplementary Fig. 5c-e). Finally, we note that the spin structure is reminiscent of other topological structures currently being researched and for chiral domain walls would represent effectively an *antiferromagnetic anti-skyrmion*. In order to ascertain how the Néel vector rotates across the line scan indicated in Supplementary Fig. 5c, we can utilize Supplementary Fig. 5d, where we plot the line scan of the contrast and is sensitive to the in-plane or out-of-plane orientation of the Néel vector. Due to the strong out-of-plane contrast of the domain wall in Supplementary Fig. 5a), we can conclude that *n* rotates out of the sample plane across the domain wall. The structure of the line scan in Supplementary Fig. 5c)-d)is schematically shown in Supplementary Fig. 5e, although whether these domain walls are chiral cannot be deduced from these measurements.

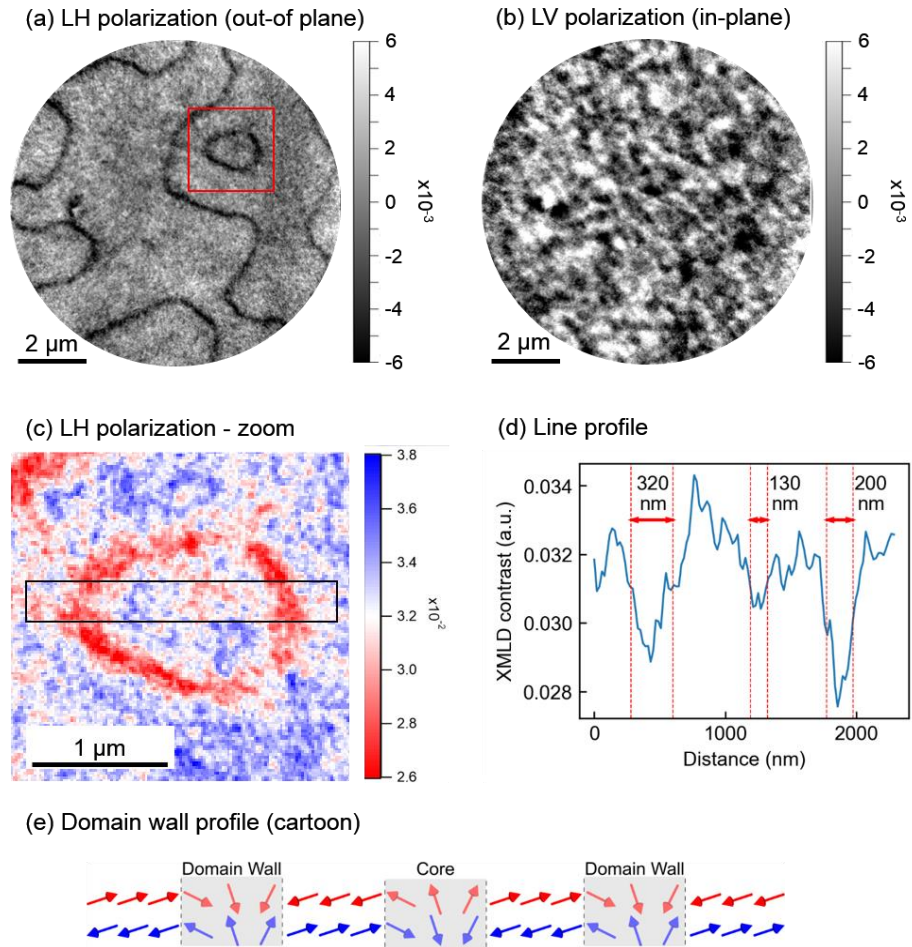

Supplementary Fig. 5. a) XMLD-PEEM image for linear horizontal polarization of the incident x-rays. Large monochromatic domains are separated by dark domains walls. b) XMLD-PEEM image for linear vertical polarization demonstrating an absence of contrast for the domain walls, indicating that the Néel vector rotates out of the plane between neighboring domains. c) Expansion of the area indicated in red in a). d) Line profile of the intensity of the structure shown in c) taken as the averaged intensity across the black box. e) Schematic of the Néel vector rotation across the structure in c). The Néel vector in each domain lies at 33° to the surface plane and rotates 180° out-of-plane between adjacent domains, and thus represents a Néel type domain wall.